\documentclass{aa}  

\usepackage{graphicx, xcolor}
\usepackage{txfonts, natbib}
\usepackage{ulem}

\newcommand{\flux}{erg\,s$^{-1}$\,cm$^{-2}$}
\newcommand{\lum}{erg\,s$^{-1}$}

\usepackage{multicol}

\begin{document}

   \title{The high-energy environment of the heavy sub-Earth GJ~367~\lowercase{b} indicates likely complete evaporation of its atmosphere}
   \titlerunning{GJ 367 b's high-energy environment}

   \author{K.~Poppenhaeger\thanks{E-mail: kpoppenhaeger@aip.de}
          \inst{1, 2}
          \and
          L.~Ketzer\inst{1,2}
          \and
          N.~Ilic\inst{1, 2}
          \and
          E.~Magaudda\inst{3}
          \and
          J.~Robrade\inst{4}
          \and
          B.~Stelzer\inst{3}
          \and
          J.H.M.M.~Schmitt\inst{4}
          \and
          P.C.~Schneider\inst{4}
          }

   \institute{Leibniz Institute for Astrophysics Potsdam (AIP), An der Sternwarte 16, 14482 Potsdam, Germany
         \and
             Institute for Physics and Astronomy, University of Potsdam, Karl-Liebknecht-Str. 24/25, 14476 Potsdam-Golm, Germany
        \and
             Institut f\"ur Astronomie und Astrophysik (IAAT), Eberhard Karls Universit\"at T\"ubingen, Sand 1, 72076 T\"ubingen, Germany
        \and
             Hamburger Sternwarte, Gojenbergsweg 112, 21029 Hamburg, Germany
             }

   \date{Accepted XXX. Received YYY; in original form ZZZ}

  \abstract
  {The planet GJ~367~b is a recently discovered high-density sub-Earth orbiting an M dwarf star. Its composition was modelled to be predominantly iron with a potential remainder of a hydrogen-helium envelope. Here we report an X-ray detection of this planet's host star for the first time, using data from the spectro-imaging X-ray telescope eROSITA onboard the Spectrum-Roentgen-Gamma (SRG) mission. We characterise the magnetic activity of the host star from the X-ray data and estimate its effects on a potential atmosphere of the planet. We find that despite the very low activity level of the host star the expected mass loss rates, both under core-powered and photoevaporative mass loss regimes, are so high that a potential primordial or outgassed atmosphere would evaporate very quickly. Since the activity level of the host star indicates that the system is several Gigayears old, it is very unlikely that the planet currently still hosts any atmosphere.}


   \maketitle

%%%%%%%%%%%%%%%%% BODY OF PAPER %%%%%%%%%%%%%%%%%%

\section{Introduction}

As exoplanet detection techniques have matured, planets in the Earth-sized regime have become detectable. Especially M dwarfs as host stars provide an advantage for the detection of small planets, since their small stellar radii yield sufficiently deep transits when a small rocky planet crosses the star. The planetary radius from transit observations can be combined with the planetary mass from radial velocity measurements to yield the planetary mean density, which in turn allows some inferences on the planetary composition. 

Exoplanets that have radii similar to or smaller than Earth have been inferred to have a range of different compositions, with Trappist-1$e$ \citep{Grimm2018} and $f$ \citep{Agol2021} having similar densities to Earth and Venus, while other exoplanets have been found to display significantly lower densities which can be indicative of water envelopes (for example, Kepler-138$b$, \citet{Jontof-Hutter2015}; Trappist-1$h$, \citet{Agol2021}). The small exoplanet GJ~367$b$ was discovered to have a surprisingly high density implying a high iron content \citep{Lam2021}. Modelling by the same authors showed that the planet can be described well if the planet is composed of iron for 90\,\% of its interior, the remaining outer $10$\,\% consisting of
 a mantle layer and a small fraction of water ice VII (a specific crystalline form of ice) as well as hydrogen and helium in gaseous form. Recently, two additional planets at longer periods were found for the GJ~367 system through radial velocity observations, and planetary parameters for planet $b$ were refined \citep{Goffo2023}.

Exoplanets have been found to undergo significant atmospheric evolution; one signpost of this is the so-called radius gap of small planets, an observed dearth of planets with radii around 2$R_\oplus$ \citep{Fulton2017, VanEylen2018}. This is typically interpreted as planets losing their primordial hydrogen-helium envelopes through some process, where bare rocky planets live below the gap and the ones with significant hydrogen-helium envelopes remaining populate the parameter space above the gap. Different processes may be driving this atmospheric mass loss; two prominent scenarios are atmospheric escape driven by the internal heat of the planet (core-driven escape, \citealt{Gupta2020}), and escape driven by the high-energy irradiation by the host star (see for example \citealt{Murray-Clay2009, Owen2017, Kubyshkina2018, Mordasini2020}), which can have a variety of different flavours depending on the assumptions made in the respective models.

Here we report on a first-time X-ray detection of GJ~367b's host star, which is an early M dwarf in the solar neighbourhood (see Table~\ref{tab:system} for an overview of the system's properties). The detection was achieved using data from the eROSITA X-ray instrument (section~\ref{data}). This allows us to determine the coronal properties of the host star (section~\ref{res}) and to characterise the high-energy environment and possible evaporation rate of the exoplanet associated with irradiation (section~\ref{disc}).

\begin{table}
\centering
\caption{Properties of the GJ~367 star-planet system from \citet{Lam2021}.}
\label{tab:system}
\begin{tabular}{ll} 
\hline\hline
Property & Value \\
\hline
\textit{stellar parameters:} &  \\
spectral type & M1.0V \\
mass & 0.454 $M_\odot$\\
radius & 0.457 $R_\odot$\\
distance &9.41 pc \\
rotation period & $48\pm2$\,d\\
[0.2cm]

\textit{planetary parameters:} &  \\
mass & 0.55 $M_\oplus$  \\
radius & 0.72 $R_\oplus$ \\
semi-major axis & 0.0071 AU \\
orbital period & 0.322 d \\

\hline
\end{tabular}
\end{table}

\section{Observations and data analysis}\label{data}

eROSITA \citep{Predehl2021} is an X-ray instrument onboard the Spectrum-Roentgen-Gamma spacecraft \citep{Sunyaev2021}. It was launched in July 2019 into an orbit around the $L_2$ Lagrange point of the Sun-Earth system. eROSITA consists of seven Wolter telescopes with one camera assembly each and is sensitive to photon energies between 0.2 and 10 keV. eROSITA started an all-sky survey in December 2019, where it scanned the whole sky every six months in great circles roughly perpendicular to the ecliptic. Any point on the sky is scanned every four hours for several eROSITA slews, with the number of slews when a given target is in the field of view depending on the ecliptic latitude of the target.

eROSITA has completed four all-sky surveys to date (named eRASS1 to eRASS4). We accessed the data around GJ~367's position in the form of photon event files for each of the four surveys, and additionally used the source catalogues produced by the consortium. Specifically, we used the stacked source catalogue in the current data reduction version from October 31 2022, as released to the eROSITA-DE consortium\footnote{The consortium-internal catalogue file name that was used was \texttt{all\_s4\_SourceCat1B\_221031\_poscorr\_mpe\_photom.fits}.}. For the photon event files we used the data processing version 020 of eROSITA-DE. 
The eROSITA source catalogue was constructed by performing a source detection on the stacked X-ray images from the first four eRASS surveys
in the 0.2-2.3 keV band, and then using the identified source positions to perform forced photometry in narrower energy bands; this stacked catalogue is referred to as eRASS:4.
For an overview of the data reduction software for eROSITA see \cite{Brunner2022}, and for an overview of the data release of the eRASS1 data see the paper by \citep{Merloni2024}.

We performed a separation-based cross-match of the eRASS:4 catalogue to the optical position of GJ~367 as identified in the Gaia\,DR3 catalogue \citep{GaiaDR3}. GJ~367 has a significant proper motion of about 0.7 arcsec per year, and it was observed by eROSITA about half-yearly from June 2020 to December 2021. Therefore we chose the star's position at the epoch of March 31 2021 as its representative mean position for the relevant eROSITA scanning period, yielding coordinates of R.A. $=$ 09:44:28.897 and Decl. $=$ -45:46:47.808. We cross-matched this optical position of GJ~367 with the eRASS:4 catalogue and found an X-ray source within an on-sky separation of 6$^{\prime\prime}$, which is within two sigma of the typical positional uncertainty of eROSITA \citep{Brunner2022}; 
we therefore identify this X-ray source with GJ~367. The effective exposure time at GJ~367's position is about $600$~s. Since eROSITA surveys scan the sky on great-circles, the effective exposure time of a sky location varies slightly for different photon energies due to energy-dependent vignetting effects, but this variation is small for the soft X-ray energies at which GJ~367 emits.

\begin{figure}
\includegraphics[width=0.55\textwidth]{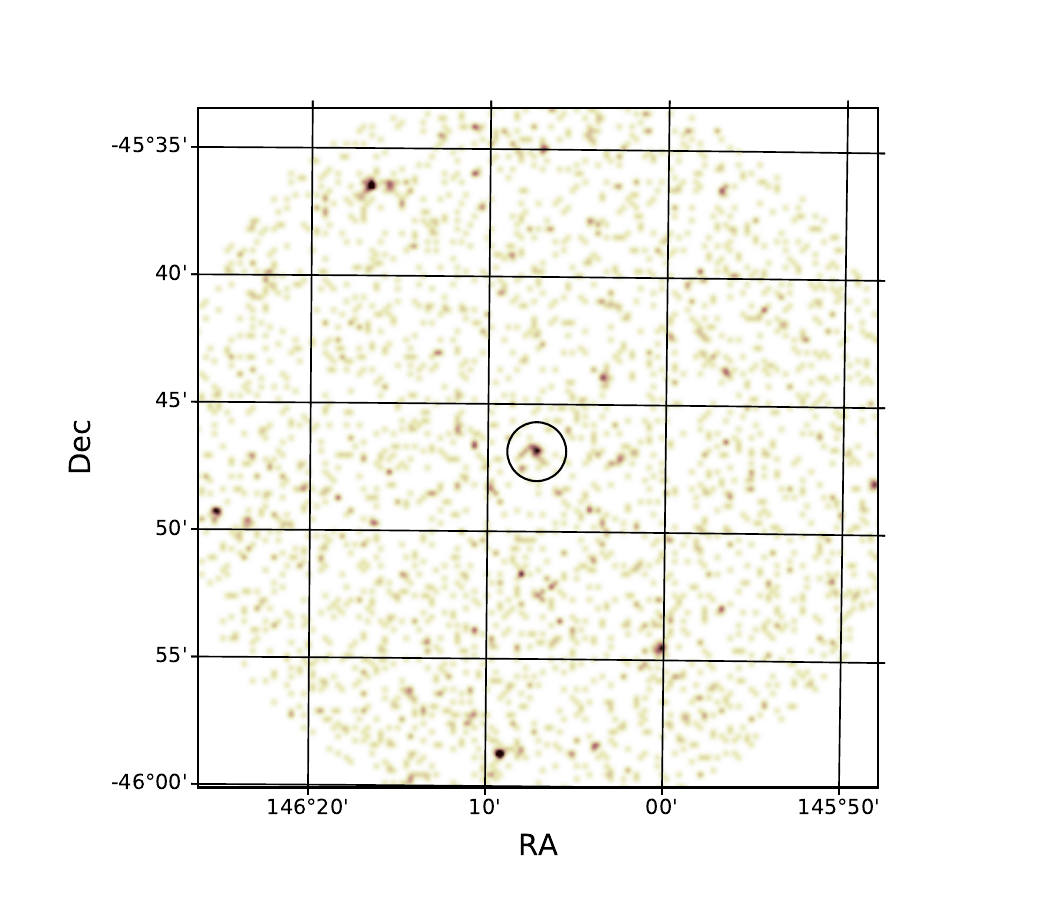}
\caption{Combined X-ray image from the first four eRASS surveys around the position of GJ~367, indicated by the circle, in the soft eROSITA X-ray band (0.2 keV to 2.3 keV). GJ~367 displays a clear excess in X-ray photons compared to the ambient background.}
\label{fig:skyview}
\end{figure}

\section{Results}\label{res}

\subsection{X-ray upper limits from previous observations}

In the past the position of the GJ~367 system was observed by the ROSAT All-Sky Survey \citep{Boller2016} and multiple times with the XMM-Newton Slew Survey \citep{Saxton2008, Freund2018}, both of which yielded unrestrictive upper limits to the X-ray flux of the star. We calculated the corresponding upper limits with the XMM-Newton Upper Limit Server\footnote{\texttt{http://xmmuls.esac.esa.int/upperlimitserver/}} to be $F_\mathrm{X,\, ROSAT} \leq 2.79\times 10^{-13}$\,\flux\, from ROSAT and the strictest of the XMM-Newton Slew Survey upper limits to be $F_\mathrm{X,\, XMM\, slew} \leq 6.33\times 10^{-13} $\,\flux. With a distance of 9.41\,pc this corresponds to an upper limit to the X-ray luminosity of the star of $L_\mathrm{X, \, ROSAT} \leq 3.0\times 10^{27}$\,\lum.

\subsection{X-ray properties from eROSITA}

The combined signal from the four completed eRASS scans of GJ~367's position clearly displays an X-ray excess; we show the combined soft X-ray image produced from the photon event files in the 0.2-2.3~keV energy band in Fig.~\ref{fig:skyview}. 
In the eRASS:4 source catalogue, the source we identified with GJ~367 is significantly detected. The detection significance is determined in the eROSITA data reduction as the negative logarithm of the probability that the observed excess counts have arisen as a chance fluctuation of the background, meaning that larger log-likelihood values correspond to more significant source detections. A minimum log-likelihood threshold of 5 was applied in the construction of the catalogue; GJ~367 has a log-likelihood value of 32 for the 0.2-2.3 keV energy band, making it a solid detection.

% see here: https://erosita.mpe.mpg.de/edr/DataAnalysis/ermldet_doc.html

\subsubsection{Stellar coronal temperature}

\begin{figure}
\includegraphics[width=0.5\textwidth]{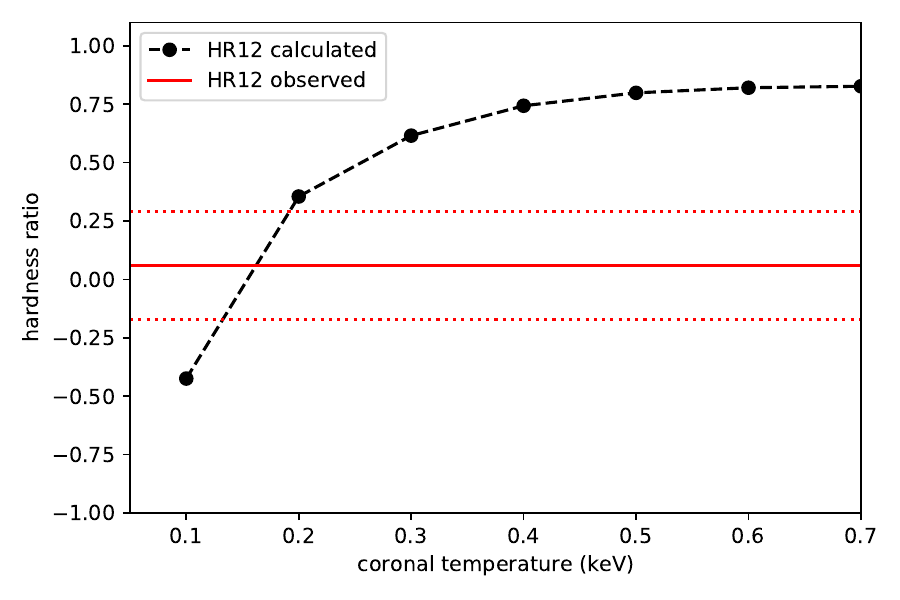}
\caption{Measured X-ray hardness ratio $\mathrm{HR}_{12}$ of GJ~367 (red solid line, with red dotted lines indicating $1\sigma$ uncertainties) compared to the same hardness ratio for simulated X-ray spectra of different coronal temperatures (black dots and dashed line). GJ 367's hardness ratio implies a mean coronal temperature in the 0.13--0.19~keV range, i.e.\ about 1.5--2.2 million K.}
\label{fig:hr}
\end{figure}

The ROSAT non-detection of GJ~367 implies that it has a moderate to low X-ray luminosity for an M dwarf, which we discuss in more detail in section~\ref{disc_1}. The star can therefore be expected to have a soft X-ray spectrum, as X-ray luminosity and temperature are correlated for stellar coronal spectra (see for example \citealt{Schmitt1997, Johnstone2015, Magaudda2022}).

Since the determination of X-ray fluxes and luminosity depends on the underlying spectral shape, we proceded by estimating the coronal temperature for GJ~367 from the collected eROSITA data.

A total of 24.9 source counts above the background 
were collected in the 0.2-2.3\,keV band; this is a maximum-likelihood estimate for the source counts that takes into account vignetting and the local background and is therefore not an integer number. This number of collected counts is small and allows only a low signal-to-noise (S/N) spectrum to be extracted. We therefore first considered the hardness ratio of the two softest bands of the star, with $HR_{12} = (R_2 - R_1)/(R_2 + R_1)$ where $R_2$ is the excess count rate in the harder band and $R_1$ in the softer band, and energy bands of 0.2-0.5 and 0.5-1.0 keV were chosen.
The hardness ratio from the catalogue photon counts is listed in Table~\ref{tab:counts} as well, with $HR_{12} = 0.06 \pm 0.23$.

To convert this into an estimate for the typical coronal temperature of GJ~367, we used the same methodology as used in other works \citep{Poppenhaeger2009, Foster2022}, i.e.\ we simulated coronal spectra over the instrumental spectral response and calculated the resulting hardness ratios as a function of coronal temperature. Specifically, we employed an optically thin thermal plasma model as appropriate for a stellar corona, and used the Xspec spectral fitting software \citep{Xspec1996} and its implemented APEC spectral model \citep{AtomDB2001, AtomDB2012} to simulate spectra on a fixed temperature grid. We show the hardness ratios for the relevant spectral bands derived from the simulated spectra in Fig.~\ref{fig:hr}, together with the measured hardness ratios for GJ~367. We found that the observed hardness ratios are in agreement with a coronal temperature in the range of $0.16\pm0.03$~keV, corresponding to 
about $1.82\pm0.34$~MK. We point out here the same caveat already mentioned by \citet{Foster2022}, that the relationship between the hardness ratio and the coronal temperature is well-defined for stellar coronae with a single dominant temperature component, but may not represent coronae well in which two or more disparate strong temperature components are present.

\begin{table}

\centering
\caption{X-ray net source count rates, hardness ratios and coronal X-ray fluxes of GJ~367 in the combined first four eRASS surveys for several energy bands.\protect\footnotemark }
\label{tab:counts}
\resizebox{\columnwidth}{!}{
\begin{tabular}{lrrr} 

\hline\hline
energy band & count rate & hardness ratio & X-ray flux  \\
& (cts\,s$^{-1}$) &  & (\flux) \\

\hline
\multicolumn{2}{l}{\textit{narrow eROSITA bands:}}  & &  \\
0.2-0.5 keV & $0.018 \pm 0.006$ & $\mathrm{HR}_{12} = 0.06 \pm 0.23$ & $(2.64 \pm 0.88) \times 10^{-14}$  \\
0.5-1.0 keV & $0.020 \pm 0.007$ & -- & $(1.57 \pm 0.53)\times 10^{-14}$\\
1.0-2.0 keV & $ \leq 0.005 $ & -- & $< 0.42\times 10^{-14}$\\
\multicolumn{2}{l}{\textit{broad eROSITA soft band:}}  & &  \\
0.2-2.3 keV & $0.041 \pm 0.010 $ & -- & $(4.71 \pm 1.08) \times 10^{-14}$ \\
\hline

\end{tabular}}

\end{table}

\footnotetext{The values are given with $1\sigma$ uncertainties; the upper limit in the 1.0-2.0\,keV band is given as the one-sided 90\% confidence range, calculated as described in \cite{Tubin2024}.}

%total 0.2-2.3: 0.041432355 +- 0.009542436
%p1 0.017659217 +- 0.005850594
%p2 0.019909054 +- 0.0066149062
%p3 0.0015963976 +- 0.0020179644
%p3 90% ulim rate: 0.00496163 cps

\subsubsection{Stellar X-ray luminosity}\label{xrayres}

The eROSITA source catalogues present source fluxes derived from count rates by assuming an underlying absorbed power-law spectrum as specified in \cite{Brunner2022}. In contrast, X-ray emission from stellar coronae is described by an optically thin thermal plasma instead of a power law, which yields different conversion factors between count rates and X-ray fluxes.

We therefore determined flux conversion factors that are appropriate for an underlying coronal spectrum with a temperature of 1.82\,MK, instead of an underlying power-law spectrum. This procedure is described in detail in \cite{Tubin2024}; in short, one simulates an absorbed power-law spectrum and the appropriate coronal spectrum over the eROSITA instrumental response and effective area, which are taken from the eROSITA Calibration Database, and calculates the resulting flux conversion factors. Using this approach, we found the following appropriate flux conversion factors ($ecf$s) for GJ~367's coronal temperature for the energy bands we are interested in, which yield the coronal fluxes when multiplying with the count rate for each band:

\begin{align}
ecf_\mathrm{\,cor,\,0.2-0.5\,keV} &= 1.49 \times 10^{-12}\,\mathrm{erg\, s^{-1}\, cm^{-2}\, count^{-1}}, \nonumber\\
ecf_\mathrm{\,cor,\,0.5-1\,keV} &= 0.79 \times 10^{-12}\,\mathrm{erg\, s^{-1}\, cm^{-2}\, count^{-1}}, \nonumber\\
ecf_\mathrm{\,cor,\,1-2\,keV} &= 0.85 \times 10^{-12}\,\mathrm{erg\, s^{-1}\, cm^{-2}\, count^{-1}}, \nonumber\\
ecf_\mathrm{\,cor,\,0.2-2.3\,keV} &= 1.14 \times 10^{-12}\,\mathrm{erg\, s^{-1}\, cm^{-2}\, count^{-1}}, \nonumber
\end{align}

%\begin{equation}
%\begin{align}
%F_\mathrm{cor,\,0.2-2.3\,keV} &= 1.22 F_\mathrm{pow,\,0.2-2.3\,keV},\\
%F_\mathrm{cor,\,0.2-0.5\,keV} &= 1.38 F_\mathrm{pow,\,0.2-0.5\,keV}, \\
%F_\mathrm{cor,\,0.5-1\,keV} &= 1.07 F_\mathrm{pow,\,0.5-1\,keV}, \\
%F_\mathrm{cor,\,1-2\,keV} &= 0.86 F_\mathrm{pow,\,1-2\,keV}, \nonumber
%\end{align}
%\end{equation}

% or as direct counts-to-flux ecfs from canned response:
% 0.2-2.3: 1.138e-12
% 0.2-0.5: 1.491e-12
% 0.5-1.0: 7.915e-13
% 1.0-2.0: 8.451e-13

%\noindent with $F_\mathrm{cor}$ being the flux for an underlying coronal spectrum with the derived temperature of 1.82\,MK, and $F_\mathrm{pow}$ being the eROSITA catalogue flux using an underlying absorbed power-law spectrum as specified in \cite{Brunner2022}. 
Applying these conversion factors yielded stellar X-ray fluxes for GJ~367 as listed in Table~\ref{tab:counts}. Unsurprisingly for such a cool corona, the emission is very soft, and within narrow energy bands, the source is only individually detected in the 0.2-0.5\,keV and 0.5-1.0\,keV band. The 1.0-2.0~keV band showed a weak excess, yielding no significant detection in that band alone. The standard eROSITA energy band of 0.2-2.3~keV yielded a total coronal flux, using the correction factor for that band, of $(4.71 \pm 1.08) \times 10^{-14}$\,\flux. We used this value for the following luminosity and surface flux calculation of GJ~367.

We derived an X-ray luminosity of $(5.0\pm1.1)\times10^{26}$\,\lum\ in the 0.2-2.3~keV energy band, using the stellar distance of 9.41~pc as listed in Table~\ref{tab:system}. The detected X-ray luminosity of the host star is, therefore, fainter than the previous ROSAT upper limit by a factor of about six.

Other relevant properties that can be calculated from the stellar X-ray luminosity are the stellar activity indicator $\log (L_\mathrm{X}/L_\mathrm{bol})$ and the X-ray flux through the stellar photospheric surface $F_\mathrm{X,\, surf} = L_\mathrm{X}/(4\pi R_{\ast}^2)$. 
We estimated the stellar bolometric luminosity from the color relations given by \citet{Mann2015,Mann2016}, which yielded $L_\mathrm{bol} = 0.030 L_\mathrm{bol,\, \odot} = 1.1\times 10^{32}$\,\lum. The activity indicator then works out to $\log (L_\mathrm{X}/L_\mathrm{bol}) = -5.34^{+0.09}_{-0.11} $ and the flux through the stellar surface to $F_\mathrm{X,\, surf} = (3.9\pm0.9)\times 10^4$\,\flux. We list all relevant X-ray quantities together in Table~\ref{tab:xrayproperties}.

% catalogue fluxes (power law): all_s4_SourceCat1B_221031_poscorr_mpe_photom.fits
%total 0.2-2.3: 3.857761e-14 +- 8.88495e-15
%P1: 1.9159397e-14 +- 6.347612e-15
%P2: 1.4649782e-14 +- 4.8674808e-15
%P3: 1.5743566e-15 +- 1.9901029e-15 i.e. not detected.

\begin{table}
\centering
\caption{Derived X-ray properties of the exoplanet host star GJ~367}
\label{tab:xrayproperties}
\resizebox{\columnwidth}{!}{
\begin{tabular}{l l}
\hline\hline
X-ray property (0.2-2.3~keV band) & Value\\
\hline
X-ray luminosity $L_\mathrm{X}$ & $(5.0\pm 1.1)\times10^{26}$ \lum \\
mean coronal temperature $T_\mathrm{cor}$& $1.82\pm0.34$~MK \\
activity indicator $\log (L_\mathrm{X}/L_\mathrm{bol})$  & $-5.34^{+0.09}_{-0.11}$ \\
X-ray flux through stellar surface $F_\mathrm{X,\, surf}$  & $(3.9\pm0.9)\times 10^4$ \flux \\
X-ray flux at planetary orbit $F_\mathrm{X,\, pl}$ & $3.5\times 10^3$ \flux \\
\hline
\end{tabular}}
\end{table}

\subsubsection{X-ray variability}

\begin{figure}
\includegraphics[width=0.48\textwidth]{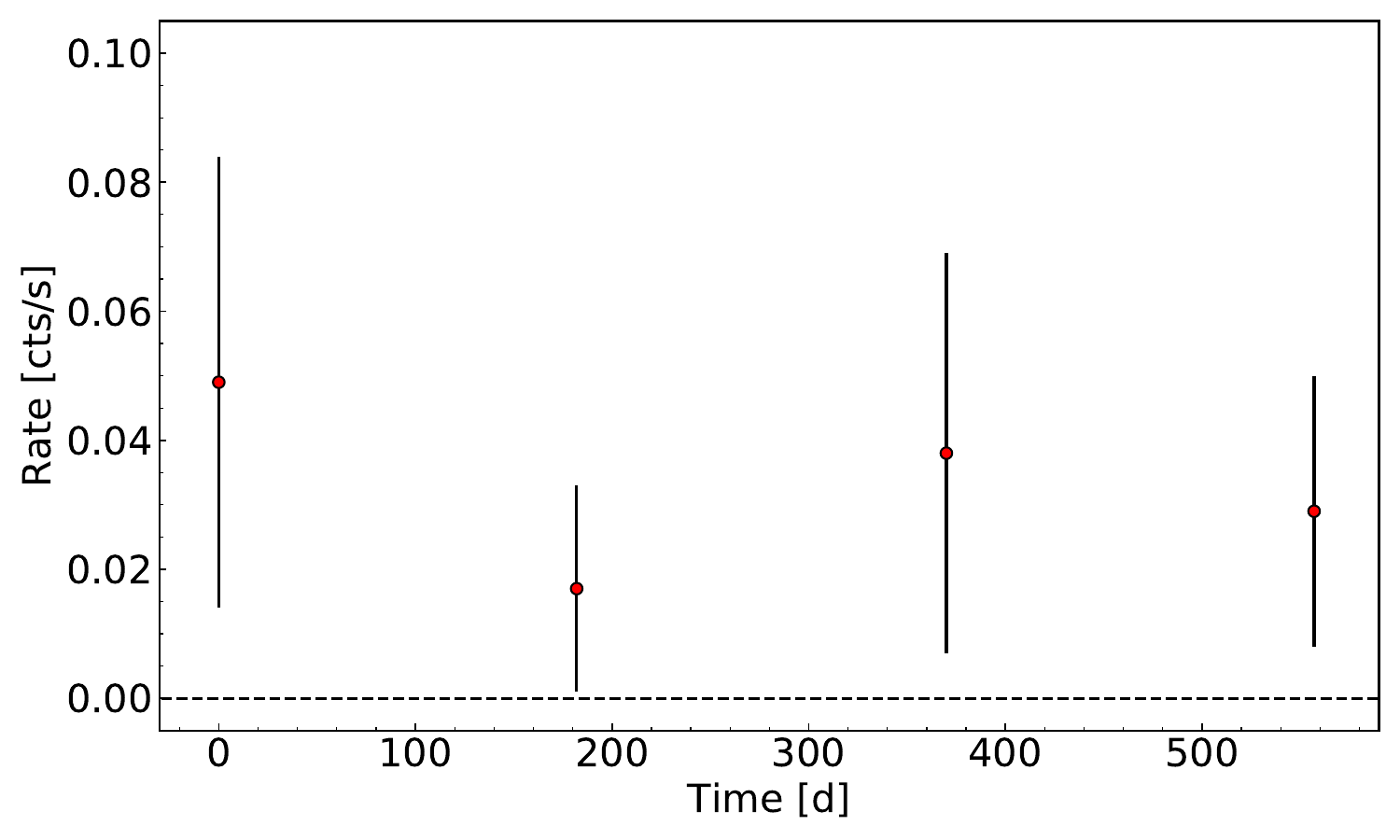}
\caption{X-ray light curve of GJ~367 over the four half-yearly eRASS surveys, displayed as source count rates in the 0.2-2.3\,keV band with $1\sigma$ uncertainties.}
\label{fig:lcspec}
\end{figure}

GJ~367 does not display any strong variability in the data collected by eRASS1:4 surveys. We show the detected mean count rates per eRASS survey in  Fig.~\ref{fig:lcspec}. They display mild variability within a factor of two, but this is consistent with a constant source flux within the uncertainties of the measurements.

\section{Discussion}\label{disc}

\subsection{The present-day magnetic activity environment of GJ~367~b}\label{disc_1}

The derived X-ray properties of GJ~367 mark it as an M dwarf with very low activity. In the context of older surveys of M dwarf X-ray luminosities, such as the sample collected by \cite{Schmitt1995} with ROSAT, it does not seem to be remarkably low in X-ray luminosity. However, this is due to the fact that GJ~367 sits at the high end of the mass and radius range of M dwarfs. The X-ray luminosity is therefore not the most suitable activity indicator to compare the star to other M dwarfs; rather, the fractional X-ray luminosity $R_\mathrm{X} = \log (L_\mathrm{X}/L_\mathrm{bol})$ is of interest, which normalises for the large size and effective temperature range present in M dwarfs of different masses. This is demonstrated by more recently analyzed X-ray samples of M dwarfs \citep{Magaudda2022, Caramazza2023}, where GJ~367 does actually sit at the lower end of observed fractional X-ray luminosities with respect to the M dwarfs that occupy  a similar mass range to GJ~367. 

\begin{figure}
\includegraphics[width=0.5\textwidth]{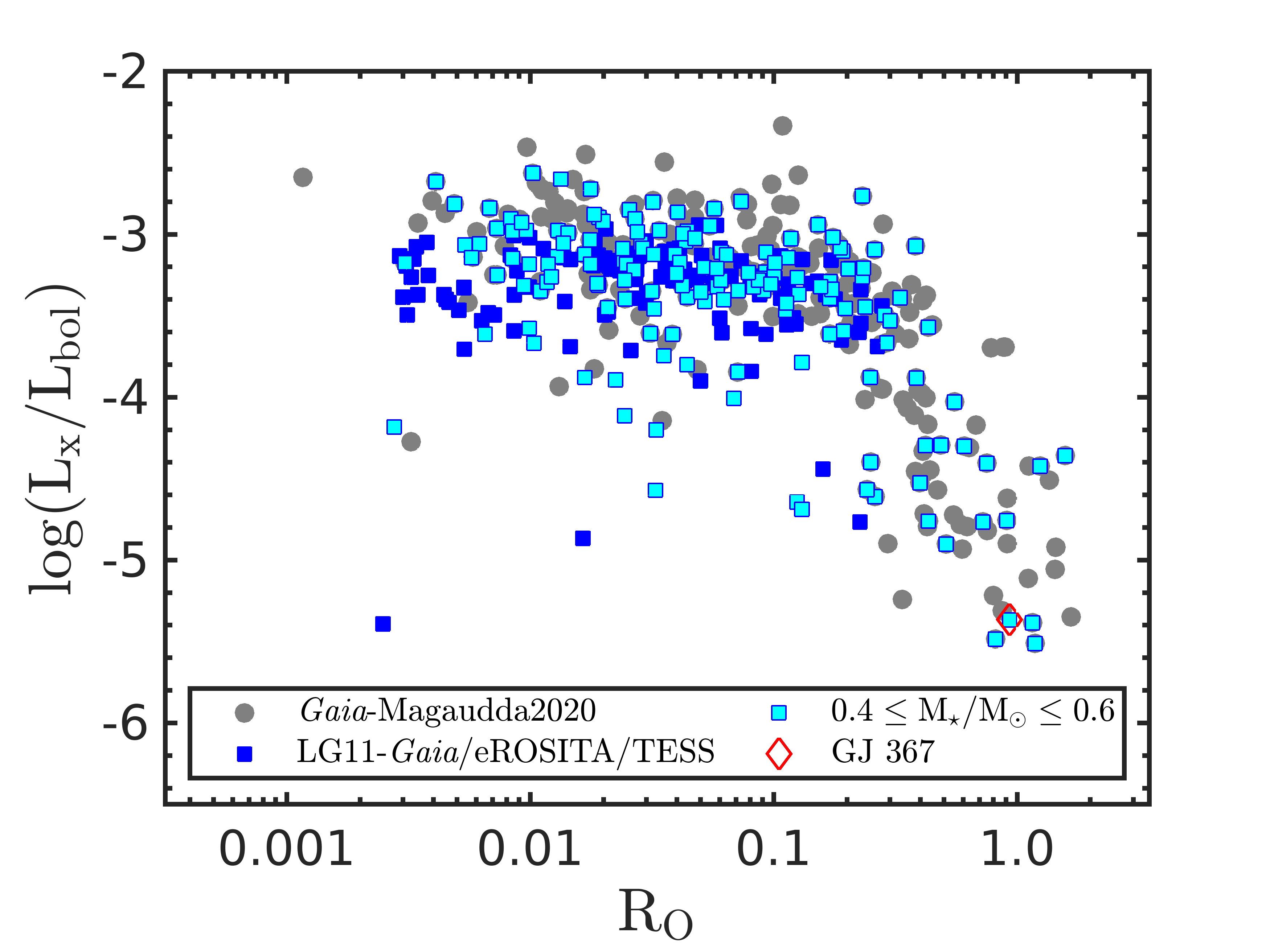}
\caption{GJ~367's X-ray activity in the context of other M dwarfs in the same mass range (from \citealt{Magaudda2022}) as a function of the Rossby number.  GJ~367 falls into the low-activity tail of the distribution and is one of the lowest-activity M dwarfs with detected X-ray emission and rotation period; however, it is not unusually X-ray inactive compared to other M dwarfs at large Rossby numbers.}
\label{fig:Lx_Lbol_Ro}
\end{figure}

The fractional X-ray luminosity $R_\mathrm{X}$ can be put into context with the stellar rotation period. All reported rotation periods for GJ~367 are quite long: \cite{Lam2021} report a period of $48\pm 2$~days from flux modulation in broad-band light curves; estimates for the rotation period derived from chromospheric activity indicators are in the range of 54 to 58 days \citep{Lam2021, Goffo2023}; and an analysis of variations in observed radial velocities yielded a period estimate of $51.30\pm0.13$~days \citep{Goffo2023}. We use here the  value of 48~days derived from light curve modulation, but we point out that using one of the other values does not make any significant difference in the following discussion. We calculated GJ~367's Rossby number $Ro = P_{\rm rot}/\tau_{\rm conv} = 0.93$ using the rotation period $P_{\rm rot}$ from \citet{Lam2021} and estimated the convective turnover time $\tau_{\rm conv}$ from the relation determined by \citet{Wright2018}. We derived the fractional X-ray luminosity  $R_X = -5.34$ in section~\ref{xrayres}. We compare this to the sample presented by \citet{Magaudda2022}, highlighting the M dwarfs with masses between $0.4$ and $0.6$\,$M_\odot$ in turquoise (see Fig.~\ref{fig:Lx_Lbol_Ro}).
Here we see that GJ~367 falls into the slowly-rotating, low-activity tail of the distribution. It is among the lowest-activity M dwarfs with measured rotation periods, but not an outlier in terms of the overall rotation-activity relation. Together with its long rotation period (see Table~\ref{tab:system}), the low X-ray activity suggests a mature main-sequence age of the star of several Gyr (see for example \citealt{Gruner2023, Curtis2020}).

% M = 0.454 # fraction of solar mass
% tau = 10**(2.33 - 1.50*M + 0.31*M**2)
% tau is 51.63 d
% Ro = 48/tau = 0.93.

While the stellar coronal temperature and absolute X-ray luminosity are on the low end compared to other M~dwarfs, the high-energy environment of the exoplanet GJ~367~b is intense nevertheless, since the planet is located at a distance of only 0.0071~AU to the host star. We calculate the high-energy flux the planet receives in its orbit from the planetary orbital parameters derived by \cite{Lam2021}, which are also listed in Table~\ref{tab:system}. We find an X-ray flux at the planetary orbit of $F_{\rm X,\,pl} = 3.5\times 10^3$~\flux{} in the 0.2-2.3 keV band. This is lower by only about a factor of two than the X-ray irradiation of the evaporating hot Jupiter HD~189733~b \citep{Poppenhaeger2013, Pillitteri2014}.

\subsection{Stellar activity effects on any remaining atmosphere of the planet}

% Lam 21: less than 1% ice and gas

Modelling of the planet GJ~367~b presented by \cite{Lam2021} suggests a core consisting predominantly of iron surrounded by a mantle, with the possibility of a thin hydrogen-helium atmosphere on top. How much of an atmosphere is to be expected from those models depends on the density of the planet. \cite{Lam2021} modelled the mass budget for the atmosphere to be up to 1\% of the planetary mass for their reported most probable density of 8.106\,g\,cm$^{-3}$, and significantly smaller for higher densities of around 10\,g\,cm$^{-3}$, as were reported by \cite{Goffo2023}. Whether very small and hot exoplanets such as GJ~367\,b actually form with a primordial hydrogen-helium envelope at all is currently unclear \citep{Neil2020, Qian2021}; however, even if no primordial hydrogen-helium envelope is present, volatiles in the magma of young rocky planets may lead to outgassing and a hydrogen-dominated atmosphere \citep{Katyal2020}. Finally, heavier outgassed atmospheres have been discussed for GJ~367\,b \citep{Zhang2024}.

{We therefore briefly discuss the stability of a potential atmosphere of GJ~367\,b under different scenarios. Two different mass loss drivers are considered to be most important for close-in exo\-planets: core-powered mass loss driven by planetary internal heat \citep{Ginzburg2018, Gupta2019}; and photoevaporation driven by stellar high-energy irradiation \citep{Watson1981, Murray-Clay2009}. An analytical approximation for photoevaporation is the energy-limited escape scenario, where the mass loss rate scales linearly with the stellar high-energy irradiation \citep{Yelle2004, Lecavelier2004, Erkaev2007}; for exoplanets with a weakly bound atmospheres, i.e.\ typically hot, close-in, and low-mass planets, hydrodynamic aspects are expected to cause much stronger mass loss rates than predicted by the energy-limited approximation \citep{Kubyshkina2018, Koskinen2022}. A useful parameter to distinguish between the different mass-loss regimes is the Jeans parameter, defined as

\begin{equation}
\Lambda = \frac{G M_\mathrm{pl} \mu_\mathrm{atm}}{k_\mathrm{B} R_\mathrm{pl} T_\mathrm{eq}},
\label{eq1}
\end{equation}

where $G$ is the gravitational constant, $M_\mathrm{pl}$ the planetary mass, $\mu_\mathrm{atm}$ the mean molecular mass of the escaping atmosphere, $k_\mathrm{B}$ the Boltzmann constant, $R_\mathrm{pl}$ the planetary radius, and $T_\mathrm{eq}$ the planetary equilibrium temperature (\citealt{Fossati2017}, see also \citealt{Guo2024}).
Photo-evaporation of atmospheres in the energy-limited regime is expected to become the dominant mass loss mechanism at large $\Lambda \gtrsim 10$ \citep{Kubyshkina2021RNAAS, FernandezFernandez2023}.

% Mpl = 0.546*5.972e27 # g
% Rpl = 0.718*6.378e8 # cm
% G = 6.674*1e-8 # cgs
% mu = (0.75*1 + 0.25*4) * 1.6736e-24 # g
% kB = 1.381e-16 # cgs
% Teq = 1597 # K 
% Lambda = G * Mpl * mu / (kB * Rpl * Teq)

% mu_co2 = 44/3 * 1.6736e-24 # g
% Lambda_co2 = G * Mpl * mu_co2 / (kB * Rpl * Teq)

For $\Lambda < 10$, i.e.\ for low-mass and hot planets, core-powered escape quickly becomes dominant. In the case of GJ~367\,b, $\Lambda = 6.3$ for a hydrogen-helium atmosphere. Typical mass loss rates in this $\Lambda$ regime have been modelled to be of the order of $10^{15} - 10^{18}$\,gs$^{-1}$ \citep{Kubyshkina2021RNAAS}. Even if we assume that GJ~367\,b currently holds a hydrogen-helium atmosphere amounting to 1\% of its planetary mass (i.e.\ as modelled by \citealt{Lam2021}), this atmospheric mass would amount to ca.\ $3\times10^{27}$\,g and would be lost within about $0.1$\,Myr even at the lower end of those typical mass loss rates. So, given the low magnetic activity of the host star and its expected age of several Gigayears, no hydrogen-helium atmosphere is expected to be currently present for GJ~367\,b.

% ; we mention as a caveat that those mass loss rates were modelled for planets with at least one Earth mass, and planets with lower mass such as GJ~367\,b can be expected to evaporate even faster.

% 0.55*5.972e27 / 1e15 / 3.154e7 

However, heavier atmospheres that may have been outgassed by the hot mantle of GJ~367\,b, such as a carbon dioxide dominated atmosphere, have been discussed in the literature as well \citep{Zhang2024}. For a carbon dioxide dominated atmosphere, GJ~367\,b's $\Lambda$ would be 53, which would place the planet in the energy-limited escape regime. We therefore estimate the X-ray and extreme-UV (together, XUV) irradiation of the planet in the following, which is the main driver of energy-limited atmospheric escape. We do point out as a caveat that the typical efficiency factors of energy-limited escape have only been modelled for hydrogen-helium atmospheres, while a much larger fraction of the incoming stellar XUV flux may drive photochemistry in heavier atmospheres and therefore not be available to drive an efficient evaporation process. So, any estimates of the energy-limited mass loss rates will be severely limited by this lacking knowledge. Following our calculation of the stellar XUV irradiation of the planet, we will therefore only briefly comment on possible escape rates.

While the X-ray luminosity is known thanks to eROSITA, the extreme-UV (EUV) emission of the host star needs to be estimated, since no observatories are currently operating in the EUV regime. EUV photons are the main contributor to the photoionization of hydrogen, which in turn can cause heating and mass loss from the planetary atmosphere \citep[e.g.][]{Watson1981, Murray-Clay2009}.

% EUV estimation
We use two empirical scaling relationships to obtain a general estimate of the EUV flux. One of them calculates the EUV flux based on an X-ray and EUV surface flux scaling relation \citep{Johnstone2021}, while the other scales X-ray and EUV energy bands for late-type stars based on synthetic XUV spectra \citep{Sanz-Forcada2011}. Both use an input X-ray band of 0.1-2.4~keV and calculate EUV fluxes for a band of 0.013 to 0.1~keV (equalling 100 to 920 \AA). We estimate GJ~367's X-ray luminosity in the 0.1-2.4~keV band to be $5.7\times 10^{26}$~\lum, using its measured luminosity in the 0.2-2.3~keV band and coronal temperature of 1.82~MK in the WebPIMMS tool\footnote{\url{https://heasarc.gsfc.nasa.gov/cgi-bin/Tools/w3pimms/w3pimms.pl}}. We then obtain $L_\mathrm{EUV} = 2.8\times10^{27}$~\lum\ and $L_\mathrm{EUV} = 6.4\times10^{27}$~\lum\ for the surface flux and synthetic spectra relations, respectively. This leads to a combined X-ray and EUV (XUV) luminosity of $L_\mathrm{XUV} = 3.3\times10^{27}$~\lum\ for the surface flux relation, and $L_\mathrm{XUV} = 7.0\times10^{27}$~\lum\ for the synthetic spectra relation. For GJ~367, the EUV flux is relatively well constrained, with both estimates agreeing within a factor of two. We do not make use of the \citet{Linsky2014} X-ray to Ly$\alpha$ relation due to its applicability to K-F spectral types, only.
Averaging between both estimates, we find an XUV irradiation flux that GJ~367\,b receives from its host star that amounts to $3.65\times 10^4$~\flux.

% Lx in 0.1-2.4 band: 1.26 times Lx in 0.2-2.3 band = 5.67e26
% surface flux 0.1-2.4 = 5.67e26/(4*np.pi*(0.457*6.955e10)**2) = 44662.8
% Fxsurf = 44662.8
% Feuvsurf_johnstone = 110*Fxsurf**(1-0.319) + 0.924*(110*Fxsurf**(1-0.319))**(1-0.0798)
% Leuv_johnstone = Feuvsurf_johnstone * 4*np.pi*(0.457*6.955e10)**2 = 2.77e27
% Lxuv_johnstone = 3.34e27
% Leuv_sanz = 10**(4.8 + 0.86*np.log10(5.67e26)) = 6.428e27
% Lxuv_sanz = 6.995e27
% Lxuv_mean = 0.5*(3.34e27 + 6.995e27) = 5.17e27
% Fxuv_mean_irrad = 5.17e27 / (4*np.pi* (0.0071*1.496e+13)**2) = 3.647e4 erg/s/cm^2.

% Evaporation rates

Keeping in mind the caveat mentioned above that efficiency factors for energy-limited escape may be very small for heavy atmospheres, we take a brief look at possible escape rates. The energy-limited mass loss rate of planets can be estimated as (\citealt{Erkaev2007}, see also \citealt{Owen2012}):

\begin{equation}
\dot{M}_{\mathrm{en-lim}} = - \epsilon \frac{(\pi R_{\mathrm{XUV}}^2) F_{\mathrm{XUV}}}{K G M_{\mathrm{pl}}/R_{\mathrm{pl}}} = - \epsilon \frac{3 \beta^2 F_{\mathrm{XUV}}}{4 G K \rho_{\mathrm{pl}}}\,.
\label{eq1}
\end{equation}
Here, $M_{\mathrm{pl}}$ and $\rho_{\mathrm{pl}}$ are the mass and density of the planet, $R_{\mathrm{pl}}$ and $R_{\mathrm{XUV}}$ the planetary radii at optical and XUV wavelengths, with $R_{\mathrm{XUV}} \approx R_{\mathrm{pl}}$ assumed for a heavy atmosphere. Impacts of Roche lobe overflow \citep{Erkaev2007} are encompassed in the factor $K$, which can take on values of 1 for no Roche lobe influence and $<\,1$ for planets filling significant fractions of their Roche lobes; in the case of GJ~367\,b, we calculate a Roche lobe radius of $2.5$ times the optical planetary radius and find a corresponding value of $K=0.43$ from the system parameters as given by \cite{Lam2021}. The high-energy XUV flux received by the planet is given by $F_{\mathrm{XUV}}$, and the efficiency of the atmospheric escape by $\epsilon$. 
The efficiency $\epsilon$ is highly uncertain; for hydrogen-helium atmospheres, they have been modelled to be on the order of 10 to 30\,\% where the energy-limited approximation applies \citep[see e.g.][]{Owen2013, Salz2016b}. Assuming \textit{ad hoc} an efficieny that is an order of magnitude lower ($\epsilon = 0.01$) for a carbon dioxide dominated atmosphere, the mass loss rate would work out to about $10^9$\,g/s. \cite{Zhang2024} determined a maximum amount of carbon dioxide dominated atmosphere of $1$~bar from a lack of an atmospheric detection in JWST observations of GJ~367\,b. This translates to an atmospheric mass of ca.\ $2.5\times 10^{21}$\,g. With the mass loss rate stipulated above, this atmosphere would be fully evaporated in less than 0.1~Myr.

As we have discussed above, the X-ray luminosity and spectral hardness of the host star suggest that the system is several Gigayears old, in line with the long rotation period derived by \citet{Lam2021}. Therefore, it is statistically very unlikely that the planet currently still hosts an atmosphere, when its density can also be modelled by an atmosphere-free iron-dominated rock. Our findings are consistent with the recently obtained upper limit on a present atmosphere of GJ~367\,b from mid-infrared JWST observations \citep{Zhang2024}.

Nevertheless, small rocky planets such as GJ~367\,b are interesting laboratories to improve our understanding of atmosphere survival. The specific case of this exoplanet shows that even low-activity M dwarfs can drive significant mass loss rates in the most compact systems.

\section{Conclusions}

We have presented the first X-ray detection of GJ~367, an M dwarf star hosting a recently discovered, extremely dense mini-Earth. We characterise the host star to have a low-temperature corona and an X-ray luminosity typical for old, inactive M dwarfs in the solar neighbourhood. Due to its proximity to the host star, a potentially present primary or outgassed atmosphere is vulnerable to mass loss. We find that a primordial hydrogen-helium atmosphere would be very quickly lost under the core-powered mass loss regime, and a heavier outgassed atmosphere can be expected to become evaporated by the stellar high-energy flux on short time scales as well. Therefore, it is unlikely that this planet currently hosts any remainders of an atmosphere.

%Given that this M dwarf's activity level is extremely low and therefore presents a comparatively mild eva\-porative environment, it can be expected that hydrogen-helium atmospheres are difficult to retain by small, short-orbit planets around M~dwarfs in general.

\section*{Acknowledgements}

The authors thank the anonymous referee for their insightful comments. KP, NI, and LK acknowledge support from the German Leibniz-Gemeinschaft under project number P67/2018. EM is supported by Deutsche Forschungsgemeinschaft under grant STE 1068/8-1. JR acknowledges support from the DLR under grant 50QR2105 and PCS  from DLR grant 50OR2102. 

This work is based on data from eROSITA, the soft X-ray instrument aboard SRG, a joint Russian-German science mission supported by the Russian Space Agency (Roskosmos), in the interests of the Russian Academy of Sciences represented by its Space Research Institute (IKI), and the Deutsches Zentrum für Luft- und Raumfahrt (DLR). The SRG spacecraft was built by Lavochkin Association (NPOL) and its subcontractors, and is operated by NPOL with support from the Max Planck Institute for Extraterrestrial Physics (MPE).
The development and construction of the eROSITA X-ray instrument was led by MPE, with contributions from the Dr. Karl Remeis Observatory Bamberg \& ECAP (FAU Erlangen-Nuernberg), the University of Hamburg Observatory, the Leibniz Institute for Astrophysics Potsdam (AIP), and the Institute for Astronomy and Astrophysics of the University of Tübingen, with the support of DLR and the Max Planck Society. The Argelander Institute for Astronomy of the University of Bonn and the Ludwig Maximilians Universität Munich also participated in the science preparation for eROSITA.

\bibliographystyle{aa}
\bibliography{katjasbib} % if your bibtex file is called example.bib
\end{document}